\documentclass[reprint,amsmath,amssymb,apsm,siunitx,twocolumn]{revtex4-2}
\usepackage{wasysym}
\usepackage{graphicx}
\usepackage{dcolumn}
\usepackage{bm}
\usepackage{capt-of}
\usepackage{float} 
\usepackage{hyperref}

\usepackage{siunitx}
\usepackage[caption=false]{subfig}
\graphicspath{ {./OptimisationImages/FocalPosition/}{./OptimisationImages/BestPeaks/} {./SimCompImages/SimExpFocalPosition/}{./IonisationInjectionImages/}{./LaserImages/}{./ShortVsLongExit/}{./V2_images}}

\begin{document}
\title{Mechanisms to control laser-plasma coupling in laser wakefield electron acceleration}

\author{L.T. Dickson$^{1}$}
\email{lewis.dickson@universite-paris-saclay.fr}
\author{C.I.D. Underwood$^2$}
\author{F. Filippi$^3$}
\author{R.J. Shalloo$^4$}
\author{J. Björklund Svensson$^5$}
\author{D. Guénot$^5$}
\author{K. Svendsen$^5$}
\author{I. Moulanier$^1$}
\author{S. Dobosz Dufrénoy$^6$}
\author{C.D. Murphy$^2$}
\author{N.C. Lopes$^7$}
\author{P.P. Rajeev$^8$}
\author{Z. Najmudin$^4$}
\author{G. Cantono$^5$}
\author{A. Persson$^5$}
\author{O. Lundh$^5$}
\author{G. Maynard$^1$}
\author{M.J.V Streeter$^{4,9}$}
\author{B. Cros$^1$}
\email{brigitte.cros@universite-paris-saclay.fr}

\affiliation{1 LPGP, CNRS, Université Paris-Saclay, 91405 Orsay, France}
\affiliation{2 Department of Physics, York Plasma Institute, University of York, York YO10 5DD, UK }
\affiliation{3 ENEA, Fusion and Technology for Nuclear Safety and Security Department (FSN), Rome, 00044, Italy}
\affiliation{4 The John Adams Institute for Accelerator Science, Imperial College London, London SW7 2AZ, UK}
\affiliation{5 Department of Physics, Lund University, SE-221 00, Sweden}
\affiliation{6 Université Paris-Saclay, CEA, CNRS, LIDYL, 91191 Gif-sur-Yvette, France}
\affiliation{7 GoLP/Instituto de Plasmas e Fusao Nuclear, Instituto Superior Tecnico, Universidade de Lisboa, Lisboa, 1049-001, Portugal}
\affiliation{8 Central Laser Facility, STFC Rutherford Appleton Laboratory, Didcot OX11 0QX, UK}
\affiliation{9 Centre for Plasma Physics, Queens University Belfast, Belfast BT7 1NN, United Kingdom }

\date{\today}

\begin{abstract}
Experimental results, supported by precise modelling, demonstrate optimisation of a plasma-based injector with intermediate laser pulse energy ($<1$ J), corresponding to a normalised vector potential $a_0 = 2.15$, using ionisation injection in a tailored plasma density profile. An increase in electron bunch quality and energy is achieved experimentally with the extension of the density downramp at the plasma exit. Optimisation of the focal position of the laser pulse in the tailored plasma density profile is shown to efficiently reduce electron bunch angular deviation, leading to a better alignment of the electron bunch with the laser axis. Single peak electron spectra are produced in a previously unexplored regime by combining an early focal position and adaptive optic control of the laser wavefront through optimising the symmetry of the pre-focal laser energy distribution. Experimental results have been validated through particle-in-cell simulations using realistic laser energy, phase distribution, and temporal envelope, allowing for accurate predictions of difficult to model parameters, such as total charge and spatial properties of the electron bunches, opening the way for more accurate modelling for the design of plasma-based accelerators.
\end{abstract}

\maketitle

\section{Introduction}
Laser wakefield acceleration (LWFA) utilises the extremely large electric fields of plasma waves produced during the interaction of intense pulsed laser light and underdense plasma to accelerate charged particles \cite{Tagima1979}. The ponderomotive force of the pulsed laser light produces a density perturbation of the electron population in the plasma over the timescale of the laser pulse. This perturbation in-turn sustains the large accelerating gradients due to the large number of displaced electrons and the short distances, on the order of the laser focal spot size.

These plasma waves produce accelerating and focusing electric fields up to three orders of magnitude greater than conventional radio frequency (rf) cavities \cite{esarey2009physics} allowing for extreme miniaturisation of the accelerating process. Whilst there has been significant progress in producing electron bunches with parameters comparable to those of classical rf linear accelerators in terms of peak energy \cite{gonsalves2019petawatt} or charge \cite{couperus2017demonstration,gotzfried2020physics}, further improvements in energy spread, divergence and stability of the electron bunch parameters - achieved simultaneously - are required for future applications such as plasma-based injectors, drivers for free-electron lasers or as particle sources for medical therapy \cite{walker2019facility,albert2016applications,chiu2004laser,svendsen2021focused,wang2021free}.

A proposed method for improving electron bunch control is the separation of injection and accelerating processes, as targeted by the EuPRAXIA design study \cite{walker2019facility}. EuPRAXIA is a European project dedicated to electron acceleration research with novel plasma-based acceleration schemes \cite{assmann2020eupraxia}. The initial stage, termed a laser-plasma injector (LPI), first produces and accelerates an electron bunch to ultra-relativistic energies before the bunch is injected into subsequent accelerating stages to achieve higher energies whilst retaining low energy spread, divergence and stable bunch pointing. EuPRAXIA targets an LPI capable of producing electron bunches with energy of \SI{150}{\mega\electronvolt}, \SI{30}{\pico\coulomb} of charge and an energy spread of 5\% \cite{walker2019facility}. Due to this dual function as source and initial accelerator, the LPI has the advantage of being very compact in comparison to rf technology required to reach similar energies. The separation of the injection and accelerating regimes would allow for each to be optimised to their respective role in the accelerator as a whole.

Self-trapping of electrons from the plasma background into the wakefield within the LPI requires high laser intensities to induce wave breaking in the non-linear regime \cite{esarey2009physics}. Reduction of the required laser intensity, and an increase in trapped charge, can be achieved through a process called ionisation injection \cite{Chen2006IonisationInjection,mcguffey2010ionization,pak2010injection}. Ionisation of the innermost electrons from dopant heavier atoms occurs only in phase with the peak laser intensity whilst the background plasma is comprised primarily of light atoms ionised at the leading edge of the laser pulse.

LWFA using ionisation-injection in structured plasma density profiles provides a large number of parameters and broad ranges for tuning electron bunch properties, such as peak energy, energy spread, charge and divergence. The stability of these parameters is key for applications such as free-electron lasers (FEL) \cite{wang2021free} and the instability of bunch energy, for example, has been correlated to laser fluctuations in previous work \cite{maier2020decoding}. Here we examine the mechanisms resulting in electron bunch deflection from the laser axis: an understudied but essential parameter for multistage acceleration \cite{thevenet2019emittance} or for high intensity QED experiments requiring precise electron bunch and secondary laser pulse alignment \cite{cole2018experimental, poder2018experimental}.

Broad electron distributions have been achieved at intermediate laser energy using ionisation-injection in gas cells \cite{audet2016investigation, liu2011all} and compared to Gaussian laser simulations; in gas jets, using several joules of laser energy \cite{irman2018improved}, the relative focal position of the laser, deep within the plasma structure, was shown to have a substantial effect on the resulting spectra due to alteration of laser-plasma coupling, and therefore, of the evolution of the laser within the plasma. In this article we discuss the effect of structuring of the plasma density downramp on electron spectra. Improvements in electron bunch energies are seen with an extension of the plasma downramp as previously predicted \cite{lee2018optimization}. Further, the evolution of peaked electron spectra with focal position are compared against simulations using realistic laser parameters to achieve accurate modelling of the resulting spectra. 

Whilst most studies of LWFA assume Gaussian laser drivers, investigations in gas jets at low \cite{nakanii2016effect} and intermediate \cite{ferri2016effect} intensity have demonstrated the effects of laser profile imperfections on accelerated electron parameters such as beam halo \cite{nakanii2016effect}, non-Gaussian laser profile and pulse phase \cite{ferri2016effect}, or spatial phase \cite{shalloo2020automation, he2015coherent}, indicating possible control of the electron bunch dynamics through laser phase and intensity distributions. Optimised configurations for electron bunch energies above 200~MeV in strongly beam loaded regimes have been achieved using a few joules of laser energy \cite{Kirchen_PRL2021} at high dopant gas concentration in a structured plasma target. 

Working at intermediate intensity and low dopant concentration we examine the effect of laser wavefront controlled by an adaptive optic (AO) on resulting electron spectra. This work provides data on the physical mechanisms to control accelerated electrons' energy spectra and alignment of the accelerated bunches with the laser axis. Experimental results are compared to particle-in-cell (PIC) simulations, and difficult to model parameters such as charge are reliably reproduced using realistic laser and plasma parameters. In this paper, numerical and experimental methods are presented in section II, followed by a discussion of results in section III which highlight the main effects on bunch quality through comparison of experimental and simulation results.

\begin{widetext}
    \begin{minipage}{\linewidth}
        \begin{figure}[H]
            \centering
            \includegraphics[trim={2cm 2cm 2cm 3cm},width=0.8\linewidth]{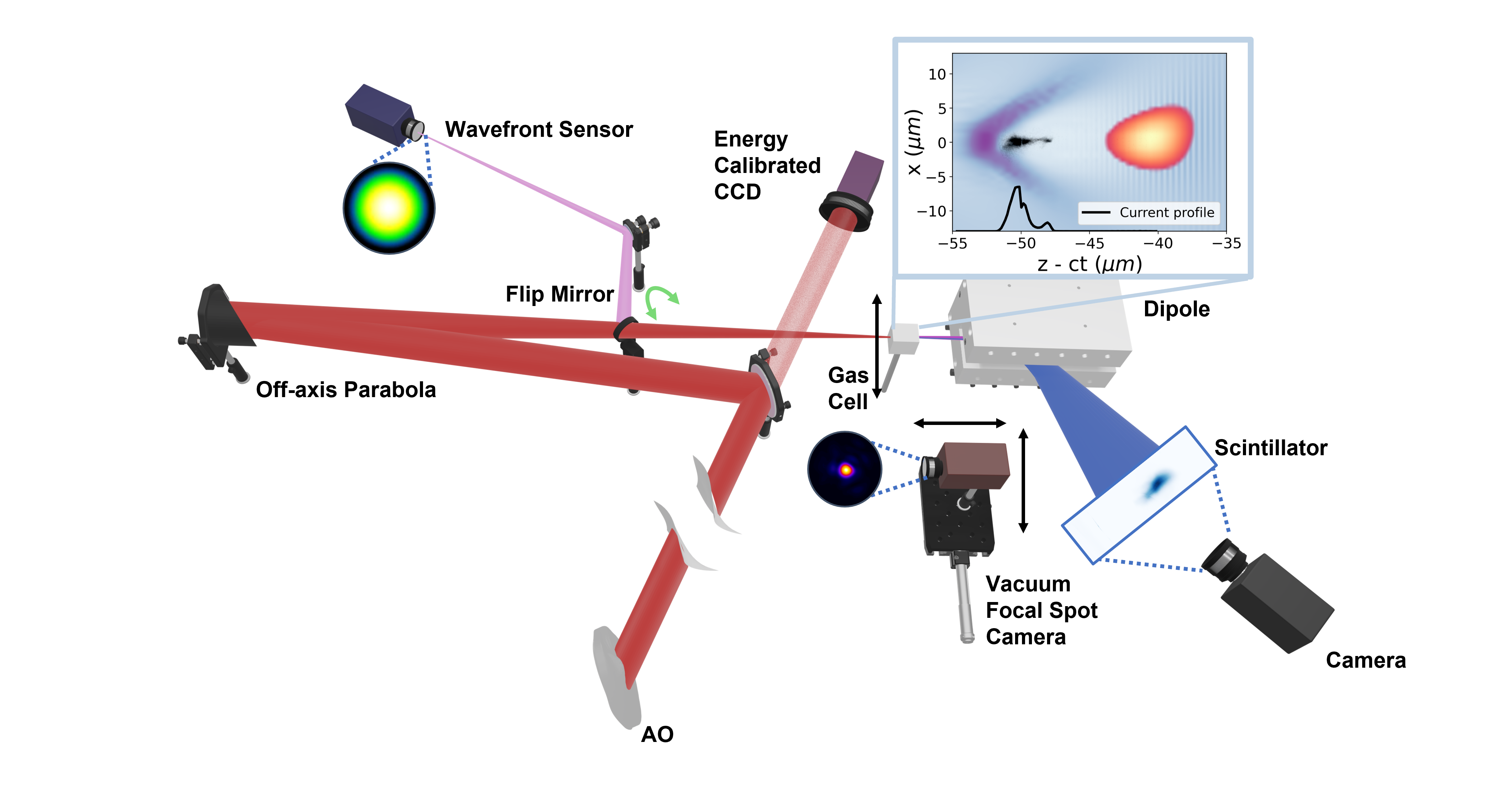}
            \caption{\label{fig:ExperimentalLayout}Experimental set up at the Lund Laser Centre. The laser is focused into the gas cell by the off-axis parabola. The interaction between the laser and the plasma produces an accelerating cavity and electron bunch which is illustrated in the simulated inset. Accelerated electrons exiting the gas cell are then dispersed with the permanent dipole magnet and produce scintillating radiation on a LANEX screen which is then imaged onto a 16-bit CCD. An Adaptive optic, set after the compressor, is used to tune the beam wavefront. Laser diagnostics are performed in vacuum using attenuators before the compressor: using the flip mirror, the beam (in pink) can be extracted to measure wavefront curvature using a Phasics wavefront sensor; the energy distribution in the focal volume is recorded in vacuum using a camera movable on axis in place of the gas cell. The adaptive optic settings are altered to produce the three laser setting cases displayed in Fig. \ref{fig:LaserFocus} as measured by the focal spot camera under vacuum. Energy measurements are taken using the leak beam (shown in light red) through a dielectric mirror and a calorimeter calibrated camera. }
        \end{figure}    
    \end{minipage}
\end{widetext}

\section{Experimental Arrangement and Methods}
\label{sec:methods}

An experiment was performed at the Lund Laser Centre (LLC) to explore laser-plasma coupling through density downramp length, focal position and laser wavefront, as means of controlling electron bunch parameters. The aim was to understand the impact of these mechanisms to approach the set of parameters, bunch energy (\SI{150}{\mega\electronvolt}), energy spread ($<$5\%) and charge (\SI{30}{\pico\coulomb}) outlined in the EuPRAXIA Conceptual Design Report \cite{assmann2020eupraxia}. As the laser pulse quality is a key component for the mechanism of ionisation injection, particular care was taken to analyse the characteristics of the laser pulse in the experiment and implement these properties in the simulation code.

An overview of the experimental arrangement is shown in Fig. \ref{fig:ExperimentalLayout}. The different aspects of the set up are described in the following subsections. The simulation result plotted in the inset shows the electron density map and laser amplitude in the xz plane (transverse to the laser polarisation plane) corresponding to the best case of this study at $z =  $\SI{1}{\milli\meter} as defined in the long exit case of Fig. \ref{fig:ShortVsLongDensity}. This inset illustrates the asymmetry of the laser shape acquired during propagation and highlights the importance of analysing the impact of the input laser mode distribution.


\subsection{Laser Pulse Characterisation and Modelling}\label{LaserProperties}

The LLC 20 terawatt laser, with an on-target energy of \SI{736}{\milli\joule} and \SI{42}{\femto\second} full-width half-maximum (FWHM) pulse duration (bi-Gaussian pulse profile \SI{25}{\femto\second} half-width at half-maximum (HWHM) before the peak and \SI{17}{\femto\second} HWHM after), and \SI{0.8}{\micro\meter} central wavelength, was focused with a $f = $ \SI{775}{\milli\meter} off-axis parabola to a focal spot size FWHM of \SI{12}{\micro\meter} achieving a peak intensity of \SI{9.8e18}{\watt\per\square\centi\meter}, corresponding to a peak normalised vector potential of $a_0 = 2.15$. Energy was measured on every shot using an energy-calibrated laser leak through the final mirror before focusing as shown in Fig. \ref{fig:ExperimentalLayout}, and an energy stability of \SI{1.95}{\percent} (std) fluctuations was measured. Tuning and control of the laser were performed in vacuum using the fully amplified laser beam, attenuated before compression to allow direct diagnostics at focus. 

\begin{figure}
    \centering
    \hspace{-0.5cm}
    \includegraphics[trim={0 0 0 0},clip, width=9cm,height=9cm,keepaspectratio]{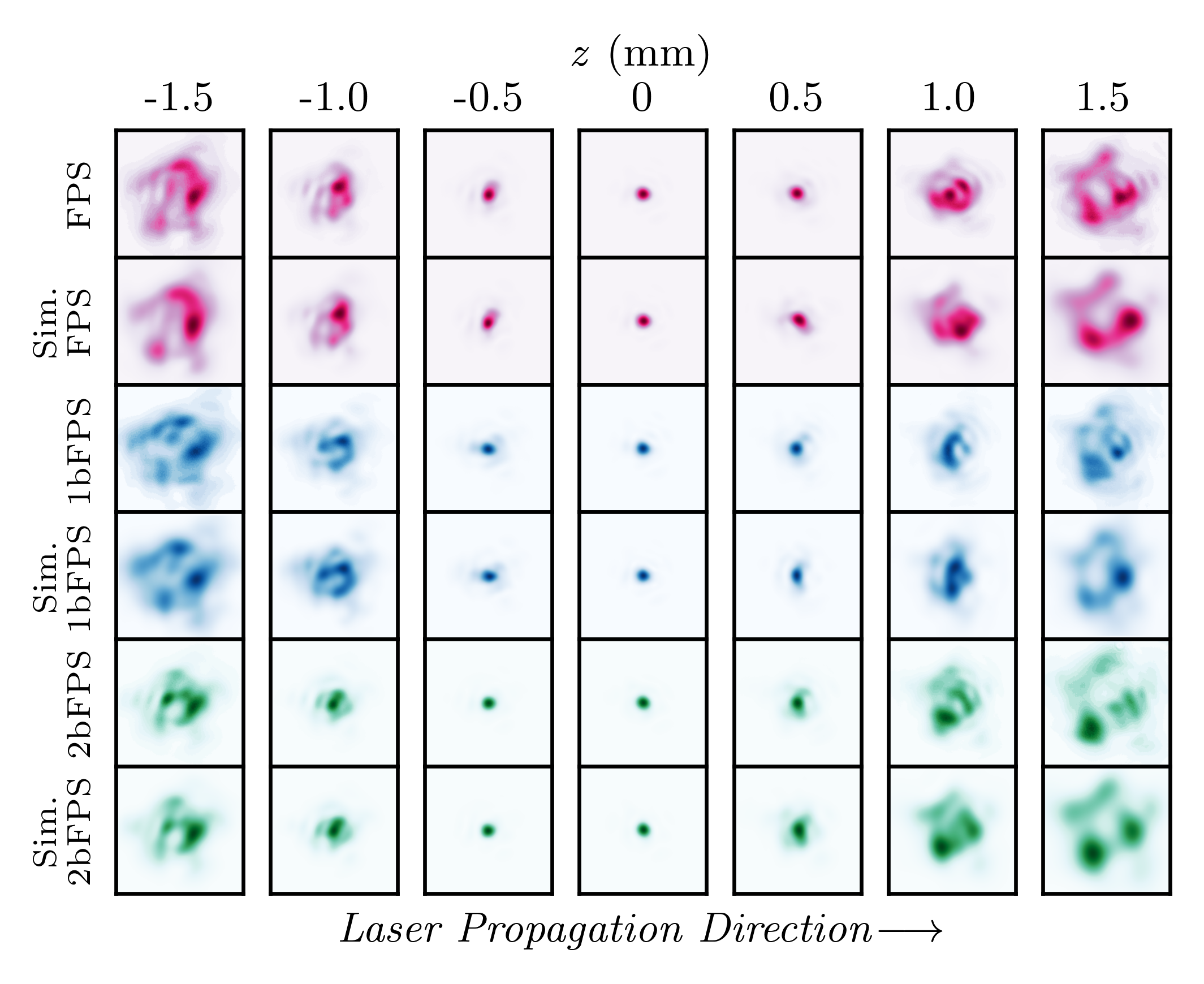}
    \captionof{figure}{\label{fig:LaserFocus} Laser energy distribution in the transverse plane around the focal position - relative position marked above - for three different settings of the AO and their corresponding energy profiles displayed from their modal description used in the simulation (denoted by \textit{Sim.}). Flat phase setting (FPS) and two manually-altered AO setting fluence profiles, 1bFPS and 2bFPS, are displayed in pink, blue and green respectively. Each image is normalised to its maximum value for visibility.}
\end{figure}

The phase front of the laser pulse was controlled with a 32 actuator NightN (opt) Ltd. brand AO \cite{kudryashov2001adaptive} in tandem with a Phasics SID4 wavefront sensor \cite{liao2006wavefront}. Images of the transverse fluence distribution, taken at different positions, $\pm1.5, \pm1, \pm0.5, 0$~\si{\milli\meter} from the focal plane along the laser axis and for three different AO settings, are shown in Fig.~\ref{fig:LaserFocus}. Each image was cropped to a \SI{130}{\micro\meter} box around their centre of mass. These fluence distributions have been obtained for three AO configurations: the wavefront sensor feedback loop provides a nearly flat phase profile at focus, FPS (flat phase settings). Next, to improve the laser pulse quality at the beginning of the laser-plasma interaction when focusing inside the plasma target, we have manually altered the AO settings to improve the pre-focal plane cylindrical symmetry at $z =  $\SI{-1}{\milli\meter}, leading to configurations 1bFPS and 2bFPS, obtained during two different experimental days. Fig.~\ref{fig:LaserFocus} shows that the three AO settings provide similar fluence distributions, particularly at the focal plane where the size of the central spot yields a Rayleigh length $z_R \simeq 400 \, \mu$m. Further, a significant variation of the laser spot shape is observed between the AO settings at each consecutive position.

Angular asymmetries can have detrimental effects on LWFA by inducing large transverse fields that can deflect the trapped electron bunch. We therefore analyse in more detail the rotational symmetry of the laser spot in a transverse plane for the three AO settings at the same longitudinal positions as in Fig. ~\ref{fig:LaserFocus}. To do so, we first define $\mathcal{R}$ as the normalised rotational asymmetry parameter (RASP). For a given transverse laser energy map $\mathcal{E}(r,\theta)$ in cylindrical coordinates, where $r$ is the radius with origin at the centre of mass and $\theta$ the azimuthal angle in the transverse plane, we define a projection $\mathcal{P}_{i}(r)$ 
for each $\theta = \theta_i$ for the laser energy map in $0 < r < R$, where $R$ is the maximum radial limit.
We then determine a rotational average $\mathcal{A}(r) = \overline{\mathcal{P}_{i}(r)\big|_{\theta=\theta_{i}}}$, $\forall i$, where each $i$ defines a single projection angle.
Finally, the RASP is obtained by calculating the normalised mean absolute variation between the rotational average and each projection:
\begin{equation}
    \mathcal{R} = \frac{\int_{0}^{R} \left(\overline{\left|\mathcal{A}(r) - \mathcal{P}_{i}(r)\right|}\right)dr}{ \int_{0}^{R}\mathcal{A}(r)dr}  \, .
\label{eq:RASP}
\end{equation}

Numerically, we take a projection every 3.6\textdegree \space around the centre of mass as a compromise between accuracy of $\mathcal{R}$ and image interpolation induced noise, which for this method was found to be of the order $10^{-5}$. From Eq.~\ref{eq:RASP} we see that the RASP for a perfect rotationally symmetric distribution (such as a Gaussian or Airy distribution) and a perfectly asymmetric one would correspond to a value of $\mathcal{R} = 0$ and $1$ respectively.

\begin{figure}
    \centering
    \includegraphics[trim={0 0 0 0},clip,width=\columnwidth]{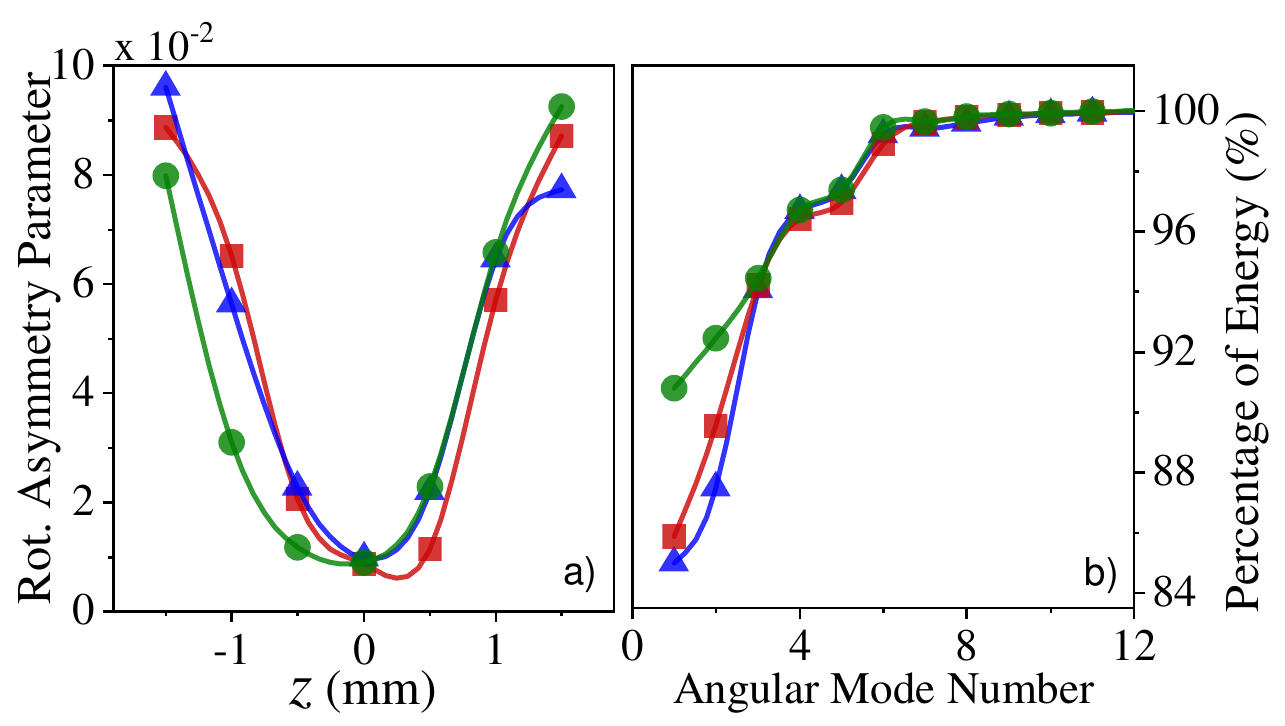}
    \captionof{figure}{\label{fig:LaserRotSymm} a) Rotational asymmetry parameter through focus for the three experimental laser energy distributions shown in Fig.~\ref{fig:LaserFocus}; b) Cumulative fraction of energy contained in successive angular modes in the simulated laser distributions. For both a), b) FPS, 1bFPS and 2bFPS are displayed in red squares, blue triangles and green circles respectively.}
\end{figure}

In Fig. \ref{fig:LaserRotSymm}a) the calculated values of $\mathcal{R}$ are plotted for the longitudinal positions and AO configurations corresponding to those in Fig.~\ref{fig:LaserFocus}. It shows that the three AO settings yield the same small minimum value of $\mathcal{R} \simeq 1\%$, obtained at the focal plane position. $\mathcal{R}$ remains close to its minimal value over a distance of $\simeq$ \SI{0.5}{\milli\meter}, then rapidly increases by more than a factor of six at $\pm$ \SI{1}{\milli\meter} from the focal position.

The experimental results for the fluence distribution reported in Fig.~\ref{fig:LaserFocus}, were used to derive an analytical form of the complex amplitude of the laser electric field (CAL). By neglecting spatio-temporal correlation, the fluence distribution can be directly transformed to an intensity distribution. In order to get the CAL from the intensity, one still needs to determine the distribution of the CAL phase. This was done using the following procedure that takes into account shot-to-shot pointing fluctuations of the laser pulse. The CAL is first projected over a large number of Hermite-Gauss (HG) functions, with a fixed origin given by the maximum intensity in the focal plane. This is done by assuming a uniform phase for the CAL in the focal plane. The three images before the focal plane ($z = -1.5$, $-1.0$, and $-0.5$ \si{\milli\meter}) are then included in a generalised Gerchberg-Saxton iteration \cite{gerchberg1972practical} to determine the phase corresponding to initially fixed origins of the HG functions at these three image positions. The positions of these origins are then determined by minimising the error between the analytical and the experimental intensity distribution at the four longitudinal positions ($z \leq  0$). Finally, the analytical intensity distributions for the other three positions ($z > 0$) are also calculated and compared to the experimental ones. As seen from Fig.~\ref{fig:LaserFocus}, the obtained analytical intensity distributions are in very good agreement with the experimental ones at all positions for the three AO settings, validating this procedure. This agreement also demonstrates the good shot-to-shot stability of the laser beam at the LLC since the phase retrieval method has converged accurately on input data ($z \leq  0$) and predicts well future positions ($z > 0$) whilst using data from different laser shots. 

From the obtained analytical expression of the CAL, one can determine the dependency of the laser energy on the azimuth angle by writing the CAL in cylindrical coordinates:

\begin{equation}
    \mathcal{A}_{\mathrm{HG}}(x,y,z,t) = \sum_{\ell= 0}^{N_\mathrm{C}-1} \mathcal{A}_{\mathrm{LG},\ell}(r,z,t)e^{j\ell \theta},
\label{eq:HG_LG}
\end{equation}
where $j^2=-1$, $\mathcal{A}_{\mathrm{HG}}$ is the CAL in Cartesian coordinates projected on HG functions while $\mathcal{A}_{\mathrm{LG},\ell}$ is the CAL corresponding to the $\ell$ angular mode, written as a sum of Laguerre-Gauss (LG) functions. $N_{\mathrm{C}}$ is the total number of complex angular modes taken into account, the total number of angular modes in real space being $N = 2N_{\mathrm{C}} -1$. The fundamental angular mode $\ell=0$ has a perfect cylindrical symmetry while the contribution of the excited angular modes $\ell \neq 0$ reflects a departure from this symmetry. 

In Fig.~\ref{fig:LaserRotSymm}b), the percentage of modal laser energy, calculated from the CAL given by Eq.~\ref{eq:HG_LG}, is plotted versus the angular mode number $N$ for the three AO settings of Fig.~\ref{fig:LaserFocus}. This plot indicates the fundamental mode contains more than 85\% of the laser energy for all the AO settings, with the 2bFPS configuration having the best cylindrical symmetry with more than 90\% of energy in the fundamental mode. The laser energy rapidly increases with $N$, 99 \% of energy being reached at $N =7$, indicating that the main part of the asymmetric contributions come from low order excited modes, and therefore justifying a description in cylindrical geometry. Moreover, this asymmetry is generated mainly in a transverse space far away from the propagation axis, where it contributes little to the plasma wave that can trap and accelerate plasma electrons. The contribution of high order excited modes is reduced when considering only the domain close to the central laser spot, in which the use of modes up to $N=5 $ already accounts for 99\% of the total laser flux. Therefore, this value of $N=5$ was used in the simulations of laser-plasma interaction presented in this article. 

To characterise the electric field of the laser at the plasma entrance, the temporal profile of the CAL envelope was determined from measurements using frequency-resolved optical gating (FROG)  \cite{trebino1997measuring} for various compressor grating separations. The optimum value of this separation (in terms of LWFA efficiency) was at the shortest pulse duration, with a FWHM pulse duration of 42 fs. These measurements also show an asymmetry between the front and back of the pulse gradients. In order to take into account this asymmetry, the pulse temporal profile was expressed using a bi-Gaussian function having a HWHM of \SI{25}{\femto\second} before the peak and \SI{17}{\femto\second} after the peak of the pulse. The spectral chirp was also extracted from this measurement and included in the simulation for completeness, although it was found to have minimal effect on the electron dynamics.

\subsection{Gas cell characteristics}
\begin{figure}[h]
    \centering
    \hspace{-0.5cm}
    \includegraphics[trim={0 0 0  0},clip,width=1.0\columnwidth]{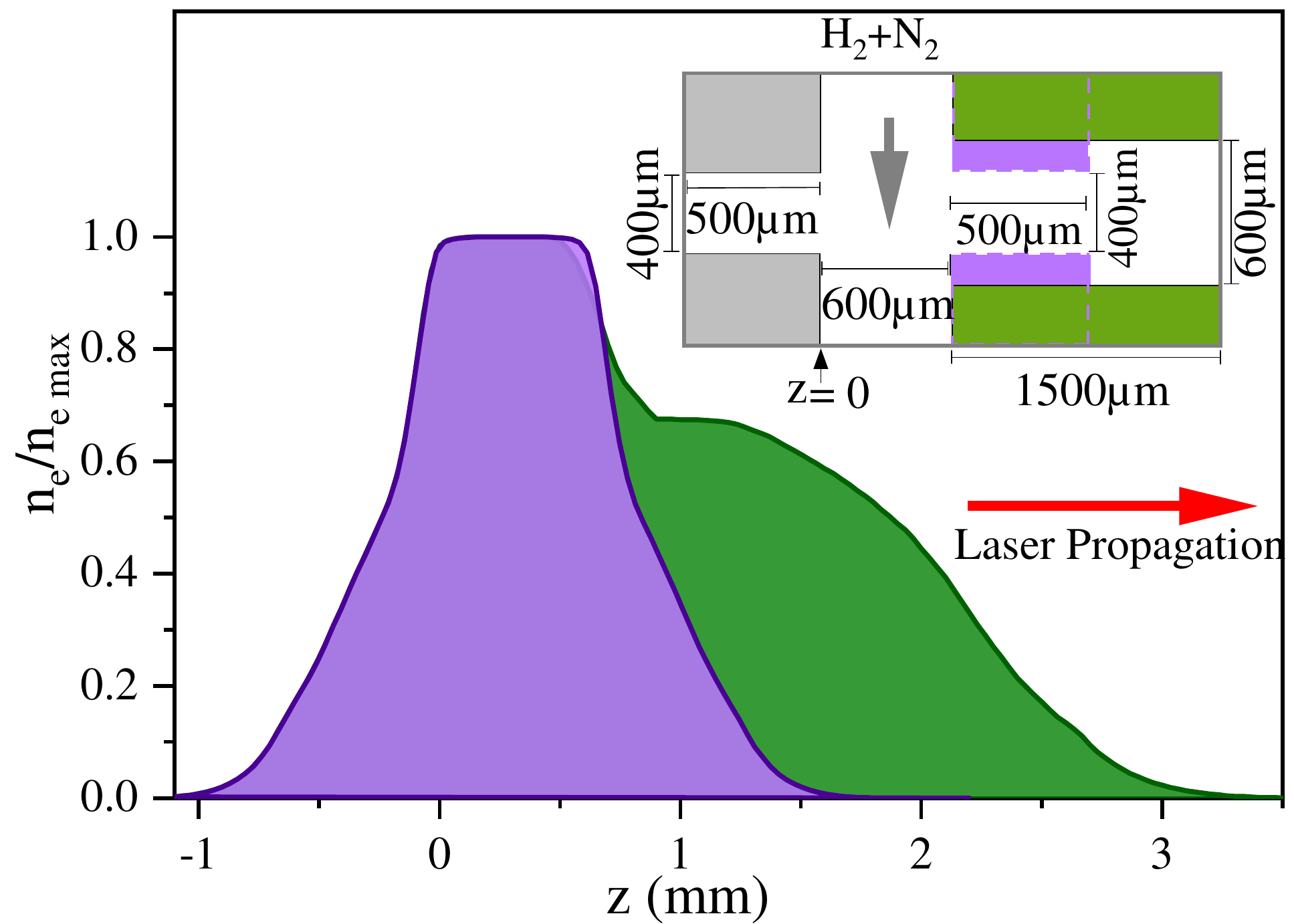}.
    \caption{Normalised electron density profile along laser axis in gas cell from fluid simulations. Short and long exit configurations are indicated by purple and green plasma densities and inset dimensions, respectively. The laser travels from negative to positive $z$ values as marked by the red arrow.}
    \label{fig:ShortVsLongDensity}
\end{figure}

Gas cell targets \cite{assmann2020eupraxia} allow for increased stability and reliability of the plasma density profile and control of the gradients for the density up and downramps which are challenging to implement in gas jets \cite{osterhoff_stable_2008,vargas2014improvements,garland2014optimisation}. Further, the process of ionisation injection must be spatially localised to limit the continuous injection of electrons throughout the plasma volume, which otherwise results in high energy spread  \cite{mirzaie2015demonstration}. To achieve localised injection in this experiment, the evolution of the laser intensity is controlled through non-linear self-focusing \cite{esarey2009physics} via a tailored plasma density implemented in the custom-built ELISA (ELectron Injector for compact Staged high energy Accelerator) gas cell used in this experiment, through variation of the aperture and length of the entrance and exit cell facings \cite{audet2016electron, lee2018optimization}. The gas cell was set in two configurations, short exit (SE) and long exit (LE), providing two different density profiles by changing the cell exit face, previously calibrated \cite{audet2016electron}, and shown in Fig.~\ref{fig:ShortVsLongDensity}.

A gas mixture of of 99.75\% hydrogen doped with 0.25\% nitrogen was chosen according to previous simulations results \cite{lee2018optimization} and following optimisation of the dopant concentration during the experimental campaign between values of 1, 0.5 and 0.25\% nitrogen. The results shown in this paper were obtained at a plasma density of n$_e \simeq 7\times 10^{18}$ cm$^{-3}$.
The gas density was calibrated off-line using a Mach–Zehnder interferometer.

\subsection{Electron diagnostics}

Measurement of the accelerated electron energy distribution was performed using an electron spectrometer composed of a \SI{20}{\centi\meter}, \SI{0.83}{\tesla} permanent dipole magnet, and LANEX scintillating screen imaged with a 16-bit Andor camera, providing an energy detection range of \SI{11.3} to \SI{300}{\mega\electronvolt} as illustrated in Fig. \ref{fig:ExperimentalLayout}. The CCD signal-to-charge calibration was performed using known intensity light sources and calibrated optical density filters along with the values by Kurz \emph{et al.} \cite{kurz2018calibration} for the count-to-charge calibration of the scintillating screen \cite{jonasThesis}.

A spatially moving mask of $\pm\SI{4}{\milli\radian}$ around the electron peak $\mathrm{dQ}/\mathrm{dE}$ value in the non-dispersive axis (vertical direction on all spectra plots) was used for both the experimental measurement and analysis of simulated results. The divergence of the moving mask was chosen to include the accelerated electron peak whilst minimising the effects of highly diffuse electrons over the measured parameters. Electron spectra are displayed within windows of $\pm\SI{7.5}{\milli\radian}$ angular width symmetrical around the laser axis; analysis of all spectra was conducted between $\pm\SI{4}{\milli\radian}$ symmetrically around the electron peak axis in the angular plane for each spectra.

Finite divergence of the electron bunch induces errors in the energy calibration since this divergence will also be present in the energy-dispersion axis. We can approximate the error in energy due to divergence through measuring the divergence in the non-dispersive axis and assuming the same divergence exists in the dispersive axis. Using this approximation, the spatial-to-energy calibration of the resulting spectra was used to calculate divergence-induced energy errors of 0.5, 1.2, and 1.6\% per milliradian divergence at 11, 150, and 300 MeV respectively. In this experiment, the laser polarisation is along the energy dispersion plane likely leading to larger divergences and therefore larger induced energy errors.  

\subsection{Simulation method}
Numerical simulations were performed with the spectral quasi-cylindrical PIC code FBPIC \cite{lehe2016FBPIC}. The complex laser amplitude at the plasma entrance was introduced through an analytical form corresponding either to a given AO setting, as described in subsection \ref{LaserProperties}, or to a Gaussian transverse profile. In the former case, the laser complex amplitude is described with $N_{\mathrm{C}} = 3$ complex angular modes ($N = 5$), while the simulation is performed with $N_{\mathrm{C}}+1$ complex angular modes to take into account the linear polarisation along the $y$-axis of the laser electric field. For a Gaussian profile, only two complex angular modes are required. In all cases, the temporal profile of the laser pulse has the bi-Gaussian form extracted from the FROG measurement. The simulations used a moving window, together with the boosted-frame technique  ($\gamma_{\mathrm{boost}} = 4$) \cite{vay2007noninvariance}. The simulation box has a dimension of \SI{70}{\micro\meter} along the propagation axis and \SI{200}{\micro\meter} in the radial direction with 2800 and 1500 cells respectively, and 48 macro-particles per cell. The initial ionisation state of plasma atoms was 1+ for H and 5+ for N. Calculations were performed at the Mesolum cluster of Université Paris-Saclay. Typical running time was 10$^4$ core-hours per simulation with four complex angular modes. Simulations took three times less in the Gaussian laser case.

\section{Results and Discussion}
Ionisation induced injection of the innermost nitrogen electrons ($N^{6/7+}$ states) is the main electron trapping mechanism at the laser intensity and plasma density used in this experiment. This has been confirmed experimentally and in the simulations, indicating minimal electron self-injection. For the value of density used ($n_{\mathrm{e}} \simeq 7\times 10^{18}$ cm$^{-3}$) self-injection requires $a_0\geq4.3$ \cite{tsung2004near}. In this experimental configuration, simulations show that $a_0$ remains below this value, even for cases leading to the highest accelerated charges. 

The maximum value of the power of the laser during the experiment was $P_0 = 16.5$~TW, and for n$_e \simeq 7\times 10^{18}$ cm$^{-3}$, the critical power $P_C$ for relativistic self-focusing is 4.2~TW. The ratio P$_0$/P$_C \simeq 4$ corresponds to the intermediate non-linear regime \cite{esarey2009physics,lu2006nonlinear}. Laser-plasma coupling during propagation, which strongly impacts electron injection and acceleration, has a strong dependency on the laser wavefront shape at the plasma entrance in this regime. It thus provides additional means to control the number of trapped electrons and the output bunch parameters. Here we analyse the relative importance of three main parameters: plasma density profile, laser focus position and laser wavefront quality on the control of the accelerated electrons as evaluated through their energy, charge and bunch angular deviation.

In the following section, we demonstrate that, in this configuration, extension of the plasma density downramp provides an increase in the electron energy and peak charge; alteration of the focal position of the laser within the plasma has a large effect on the total trapped charge and displacement from the laser axis for the accelerated bunches; and finally, alteration of the laser symmetry can be used to improve the accelerated electrons in terms of divergence and energy spread, down to the \si{\milli\radian} and per cent level, respectively, whilst minimising the amount of charge in the low energy part of the spectra.

\subsection{Plasma exit gradient}
\label{sec:downrampmanipulation}

Simulations of ionisation-induced injection in a laser-driven plasma wakefield \cite{Lee2016,lee2018optimization} show that high-quality electron injectors in the 50–200~MeV range can be achieved in a gas cell with a tailored density profile. Extending the plasma exit downramp was shown numerically to provide an increase in peak and maximum electron energy of the accelerated bunches. This effect was observed experimentally and is illustrated in Fig.~\ref{fig:earlyfocus_shortvslong} and ~\ref{fig:plateaufocus_shortvslong}. AO settings correspond to the FPS case with laser focus in the up-ramp at $z=$\SI{-0.35}{\milli\meter} for Fig.~\ref{fig:earlyfocus_shortvslong} and at the beginning of the density plateau, $z=$\SI{0}{\milli\meter}, for Fig.~\ref{fig:plateaufocus_shortvslong}. For each case of focus position, experimental electron spectral density images in the angular-energy plane illustrate a) short exit (SE) and b) long exit (LE) configurations. For all $\mathrm{dQ}/\mathrm{dE}$ plots the solid line corresponds to the spectral density images displayed to their left with the standard deviation of multiple consecutive shots.

\begin{center}
\hspace{-2.4cm}
\includegraphics[trim={0 0 2.5cm 0},width=9cm,height=9cm,keepaspectratio,width=0.8\columnwidth]{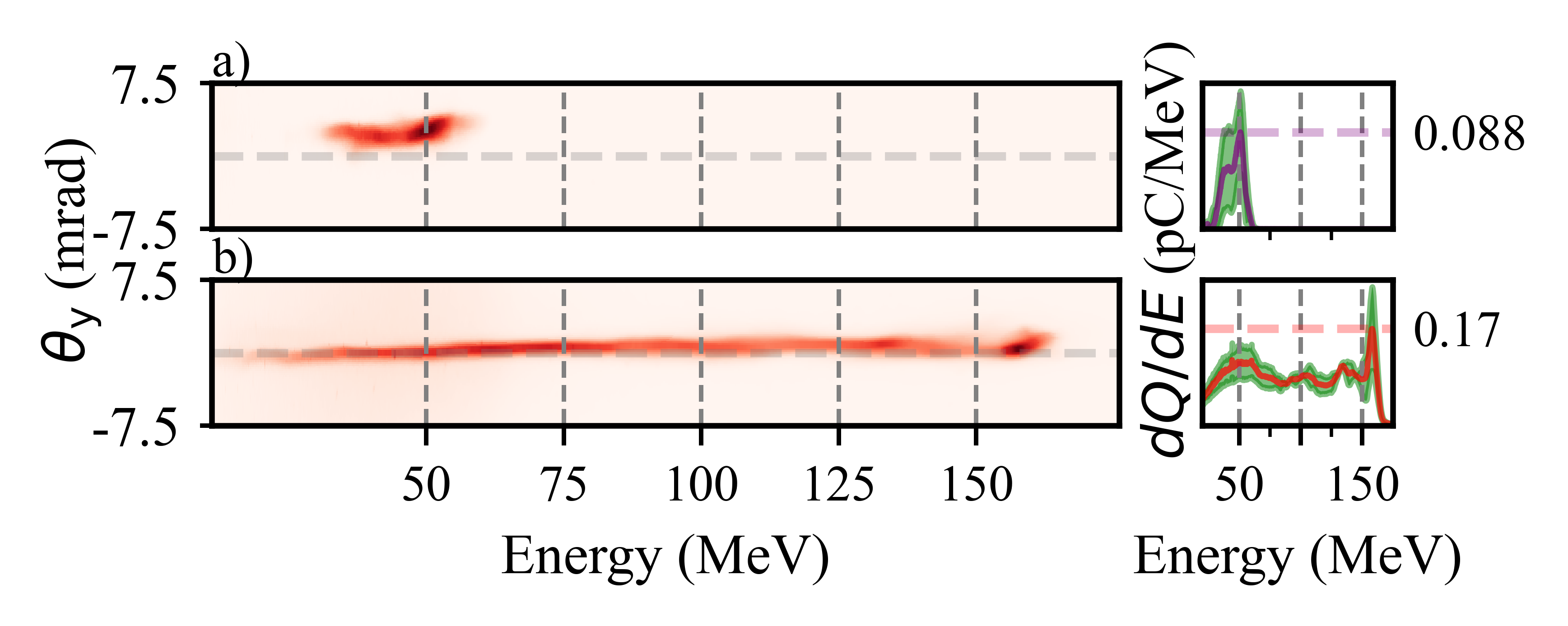}
\captionof{figure}{\label{fig:earlyfocus_shortvslong} Experimental electron charge density in divergence-energy space and their corresponding spatially-integrated $\mathrm{dQ}/\mathrm{dE}$ (\si{\pico\coulomb\per\mega\electronvolt}) within a $\pm$\SI{7.5}{\milli\radian} window around the laser axis indicated by the dashed horizontal grey line with the laser focus at $z=$\SI{-0.35}{\milli\meter} for the two exit plate configurations illustrated
in Fig. \ref{fig:ShortVsLongDensity}: a) short exit configuration (SE) case, and b) long exit configuration (LE). Standard deviation of four and three consecutive shots for a) and b) respectively are illustrated by the shaded green region. All $\mathrm{dQ}/\mathrm{dE}$ plots are plotted from zero (\si{\pico\coulomb\per\mega\electronvolt}) in linear scaling. The purple and red dashed lines indicate the maximum value of $\mathrm{dQ}/\mathrm{dE}$ for for the displayed spectra.}
\end{center}

\begin{center}
\hspace{-2.4cm}
\includegraphics[trim={0 0 2.5cm 0},width=9cm,height=9cm,keepaspectratio,width=0.8\columnwidth]{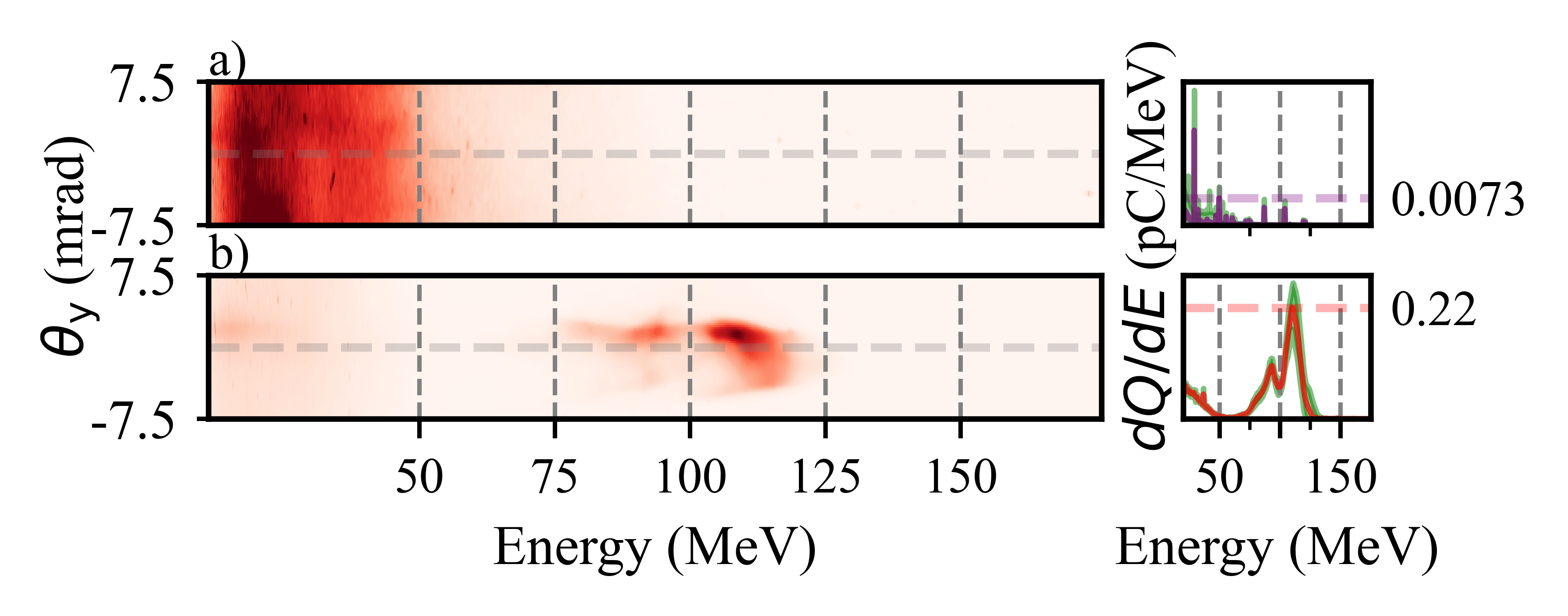}
\captionof{figure}{\label{fig:plateaufocus_shortvslong} Same as Fig. \ref{fig:earlyfocus_shortvslong} for laser focus at $z=\SI{0}{\milli\meter}$ with three consecutive shots included in the standard deviation.}
\end{center}

For both laser focus positions, measured electron spectra show an increase of almost an order of magnitude in spatial integrated charge density and approximately \SI{100}{\mega\electronvolt} in maximum energy in the LE case, compared to the SE case. Extending the density downramp from \SI{500}{\micro\meter} to \SI{1500}{\micro\meter}, corresponding to the change in plasma structure illustrated in Fig.~\ref{fig:ShortVsLongDensity}, increased the peak energy from \SI[separate-uncertainty=true]{51(2)}{\mega\electronvolt} to \SI[separate-uncertainty=true]{158(11)}{\mega\electronvolt}, as illustrated in Fig.~\ref{fig:earlyfocus_shortvslong}. This corresponds to an average accelerating gradient greater than \SI{100}{\giga\electronvolt\per\meter} throughout the density downramp.

Comparing peak $\mathrm{dQ}/\mathrm{dE}$ values from Fig.~\ref{fig:earlyfocus_shortvslong}a) and Fig.~\ref{fig:earlyfocus_shortvslong}b) demonstrates an increase in the accelerated charge-energy density of 1.9 times for the LE case, indicating that trapping continues to occur in the plasma downramp region, leading to a broad energy spectrum. Reduction in energy spread of the electron bunches is achieved for the LE case by moving the focus position of the laser to the beginning of the plasma density plateau, as illustrated in Fig.~\ref{fig:plateaufocus_shortvslong}b). A broad, low energy spectrum is produced when the same settings are used in the SE case (Fig. \ref{fig:plateaufocus_shortvslong}a)), again demonstrating that the elongated density downramp plays a key role in the injection and acceleration process.

The LE configuration was used for all results shown in the following sections.

\subsection{Optimisation of Laser-Plasma Coupling Through Focus Position}
\label{sec:optimsationfocalposition}
The position of laser focus relative to the density profile is one of the main input parameters that can be used to tune the electron bunch properties. The focal position defines the initial conditions for laser-plasma coupling through self-focusing therefore changing the resulting accelerating fields and electron bunch dynamics. In addition to the density downramp increase of the LE case, further control and improvements of the electron spectra were achieved by exploring the focal position of the laser with respect to the plasma density profile. Guiding simulations for this campaign predicted improvements in accelerated electron parameters by focusing close to $z = 0$.

\begin{figure}[h]
    \renewcommand{\thesubfigure}
    \subfloat{%
        \includegraphics[trim={0 0 0 0},clip,clip,width=\columnwidth]{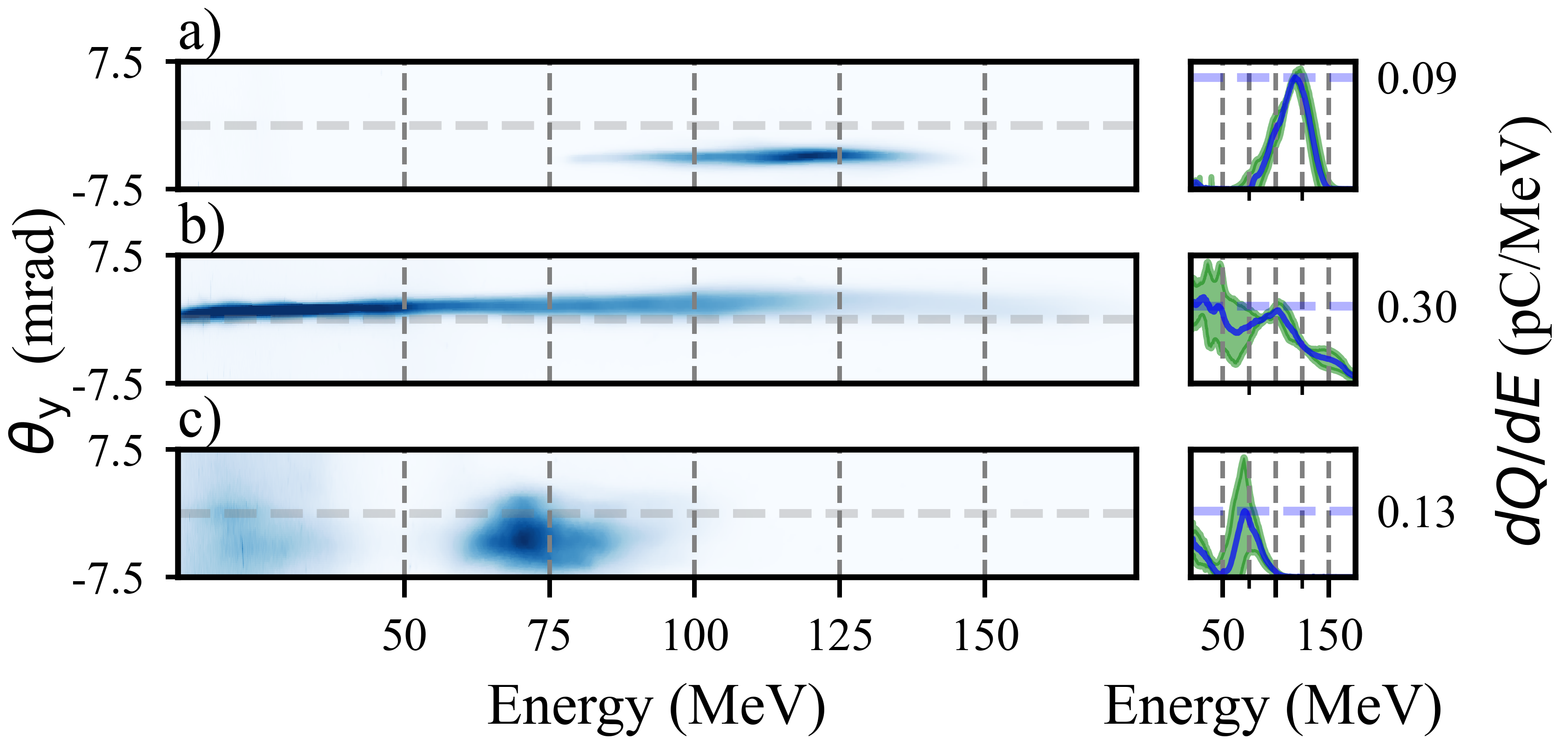}%
    }
    \renewcommand{\thesubfigure}
    \subfloat{%
        \includegraphics[,width=0.9\columnwidth]{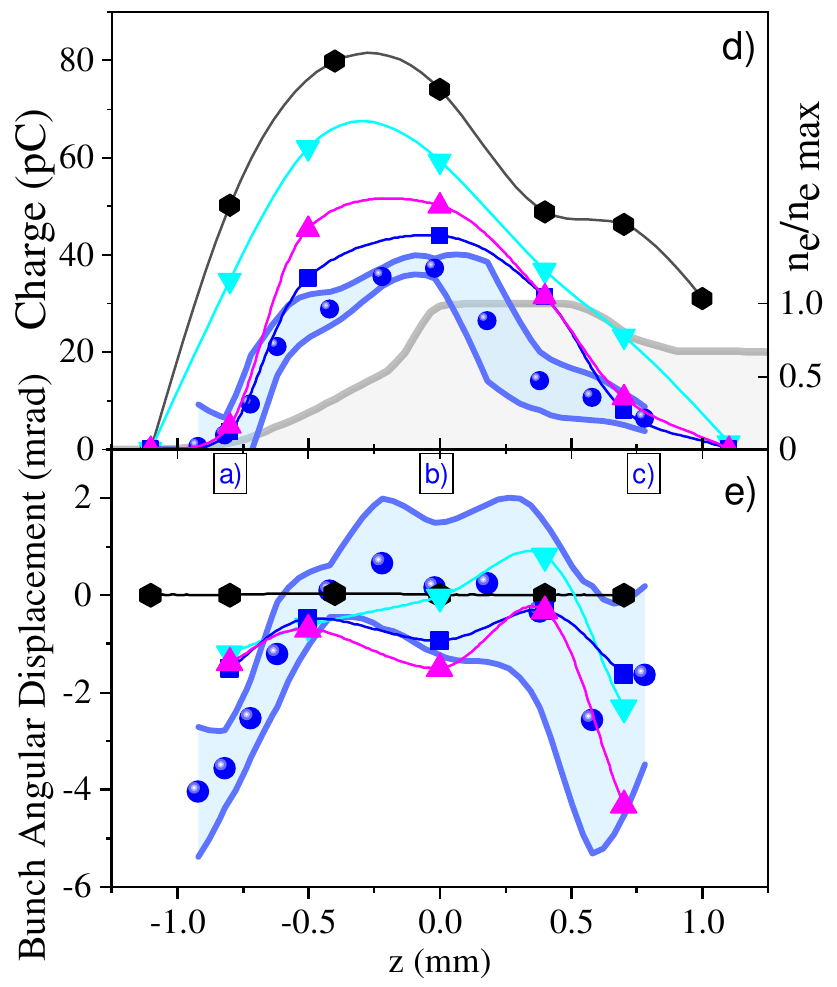}%
    }
\caption{\label{fig:Fig6_combined}Representative experimental electron spectra within $\pm$\SI{7.5}{\milli\radian} window around the laser axis between 11 and \SI{175}{\mega\electronvolt}, corresponding to laser focal positions along the longitudinal spatial axis, z, a)~pre-plateau: $z = $ \SI{-0.8}{\milli\meter}, b)~peri-plateau: $z = $ \SI{0}{\milli\meter}, and c)~post-plateau: $z = $ \SI{0.8}{\milli\meter} for the LE case and 1bFPS AO settings; d)~total charge within a $\pm$\SI{4}{\milli\radian} of the electron bunch peak spatial location, and e) electron bunch peak displacement from laser axis, defined as zero angular displacement, are shown as blue circles as functions of position along the laser axis. Simulated results for n$_e = 6.7\times 10^{18}$ cm$^{-3}$ and $a_0 = 2.15$ are plotted as black hexagons for a Gaussian laser bunch and realistic 1bFPS sim laser as blue squares; for comparison n$_e = 7.5 \times 10^{18}$ cm$^{-3}$ and $a_0 = 2.15$ are indicated by cyan down triangles ($a_0 = 2.0$ by magenta up triangles). Plasma density profile is illustrated by the grey line. Errors are given by the standard deviation of the values for both parameters and $\mathrm{dQ}/\mathrm{dE}$ from consecutive shots.}
\end{figure}

Fig.~ \ref{fig:Fig6_combined}~a) to c) show representative experimental electron spectral density images in the angular dispersion-energy plane at three laser focal positions, a)~pre-plateau: $z=$\SI{-0.8}{\milli\meter}, b)~peri-plateau: $z=$\SI{0}{\milli\meter}, and c)~post-plateau: $z=$\SI{0.8}{\milli\meter} for the LE case and 1bFPS AO settings. The average total charge and vertical displacement of the electron spectra over multiple shots are plotted at different laser focus positions relative to the plasma density distribution (indicated by the grey line in Fig.~\ref{fig:Fig6_combined}d)). Total charge for the spectra are calculated inside a mask of $\pm$\SI{4}{\milli\radian} of the electron bunch peak spatial location. The bunch peak divergence of the electron spectra are calculated from the laser axis to the peak of the spectra in the spatial dimension. The light blue shaded area indicates the amplitude of shot-to-shot fluctuations.

Fig.~\ref{fig:Fig6_combined}d) shows that the accelerated charge is strongly dependent on the focal position of the laser with a characteristic length $\simeq$ \SI{0.5}{\milli\meter} close to the value of the Rayleigh length z$_R$. This result is in accordance with the variation of the laser fluence profiles with the focal position shown in Fig.~\ref{fig:LaserFocus}. As relativistic self-focusing becomes efficient slightly before the density plateau, when the focal plane is too far from this position the laser intensity cannot reach high enough values for trapping a significant amount of charge (as seen in Fig. \ref{fig:Fig6_combined}d)) and the large deformation of the laser radial profile leads to a large bunch angular deviation (as seen in Fig. \ref{fig:Fig6_combined}e)).

Focusing at $z=$\SI{-0.8}{\milli\meter} produces a spectrum peaked at \SI[separate-uncertainty=true]{118(5)}{\mega\electronvolt} with an energy spread of \SI{27}{\percent}, a divergence of \SI[separate-uncertainty=true]{2.9(6)}{\milli\radian} (full angle), and an average bunch deflection of \SI[separate-uncertainty=true]{-3.6(3)}{\milli\radian}. In this case, the accelerated charge is low at \SI[separate-uncertainty=true]{3.32(64)}{\pico\coulomb} due to the reduced effect of self-focusing because the laser starts diverging before self-focusing. However, the percentage of charge within $2\times$~FWHM of the peak reaches \SI{94.8}{\percent} showing that the majority of accelerated electrons are within the peak, leading to an exceptionally clean signal.

Increasing the focal position to $z = 0$ produces spectra with the highest charge (Fig. \ref{fig:Fig6_combined}b) with peak $\mathrm{dQ}/\mathrm{dE}$ of \SI[separate-uncertainty=true]{0.30(6)}{\pico\coulomb\per\mega\electronvolt} at \SI[separate-uncertainty=true]{102(4)}{\mega\electronvolt} and charge of \SI[separate-uncertainty=true]{33.6(66)}{\pico\coulomb}. These bunches have improved coaxiality with the laser axis with a reduced average displacement close to zero for these settings and a slightly increased divergence of \SI[separate-uncertainty=true]{4.4(6)}{\milli\radian}. The reduced value of the peak energy and the broad energy spectra that extend up to \SI[separate-uncertainty=true]{200(14)}{\mega\electronvolt} give a signature that a large plasma-wave accelerating field is generated, but it is significantly reduced for a large part of the trapped electrons by beam loading effects. 

Finally increasing the focal position to $z =$~\SI{0.8}{\milli\meter} decreases the trapped charge down to \SI[separate-uncertainty=true]{10.7(21)}{\pico\coulomb} with a peak energy of \SI[separate-uncertainty=true]{71(10)}{\mega\electronvolt} and an energy spread of \SI{39}{\percent} (spectrum Fig. \ref{fig:Fig6_combined}c)). The presence of low energy electrons indicates two different zones of trapping. The bunch broadens with a \SI[separate-uncertainty=true]{13.0(6)}{\milli\radian} full angle divergence. Focusing at $z =$~\SI{0.8}{\milli\meter} further increases the fluctuations in the electron bunch pointing as seen by the increases in the errors due to the increased sensitivity to the laser energy distribution pre-focus. More generally, comparing \ref{fig:Fig6_combined}d) with \ref{fig:Fig6_combined}e), larger fluctuations in the bunch angular deviation than in the total charge is observed.

Simulations were performed for different settings of the input laser pulse; the resulting electron charge and bunch angular deviation are plotted in Fig. \ref{fig:Fig6_combined}d) and e) for comparison with experimental data. Gaussian pulse case (black hexagons) is compared to a realistic transverse distribution using 1bFPS settings (plotted as blue squares), for $a_0=2.15$ at n$_e = 6.7 \times 10^{18}$ cm$^{-3}$, which are the estimated experimental values. In order to show the influence of the plasma density and the laser intensity, we have also plotted simulation results for n$_e = 7.5 \times 10^{18}$ cm$^{-3}$ with either $a_0=2.15$ (cyan down triangles) or $a_0=2.0$ (magenta up triangles).

The overall dependence of charge against focal position is adequately described in the four simulation cases. However, the Gaussian calculation results in an overestimated charge, with an error of more than a factor of two close the maximum charge, and significantly higher simulated charges for early and late values of focal position. Meanwhile, simulated results using realistic laser parameters provide good agreement to experimental results: the fast decrease of the charge at late focal positions is well reproduced. Further, the overestimation of the maximum charge is only 17\% in the realistic case. Increasing the density in the simulation by 12 \% leads to an additional increase of 41\% for the value of the maximum of charge and a broadening of the corresponding charge curve in Fig.~\ref{fig:Fig6_combined}d) (cyan down-triangles), which approaches the Gaussian case. Finally, as seen in Fig.~\ref{fig:Fig6_combined}d) (magenta up-triangles), a decrease of 14 \% of the laser energy compensates the effect of the density increase at early and late focal positions.

As expected, the electron bunch remains aligned with the laser axis for all focus positions when the axis-symmetric Gaussian pulse is used (see Fig.~\ref{fig:Fig6_combined}e)). Experimental data for the angular deviation of the bunches are well reproduced by the simulation when including the realistic laser complex amplitude. In particular, the counter-intuitive fact that the sign of the displacement is unchanged when going from large negative $z$ values to large positive ones. In the former case, laser-plasma interaction occurs mainly in front of the focal plane, whereas in the latter case, it is behind. Between these two positions, there is a change of sign of the laser phase in vacuum, but not for the electron displacement inside the plasma, indicating that large non-linear effects determine the final direction of the electron bunch. Fig. \ref{fig:Fig6_combined}e) shows that the bunch angular deviation exhibits similar trends as the charge along the laser axis for variations of 12 \% in plasma density or 14 \% in laser power. 

\begin{figure}
    \centering
\hspace{-1.4cm}
\includegraphics[trim={1cm 0 2.8cm 0},width=6.5cm,keepaspectratio]{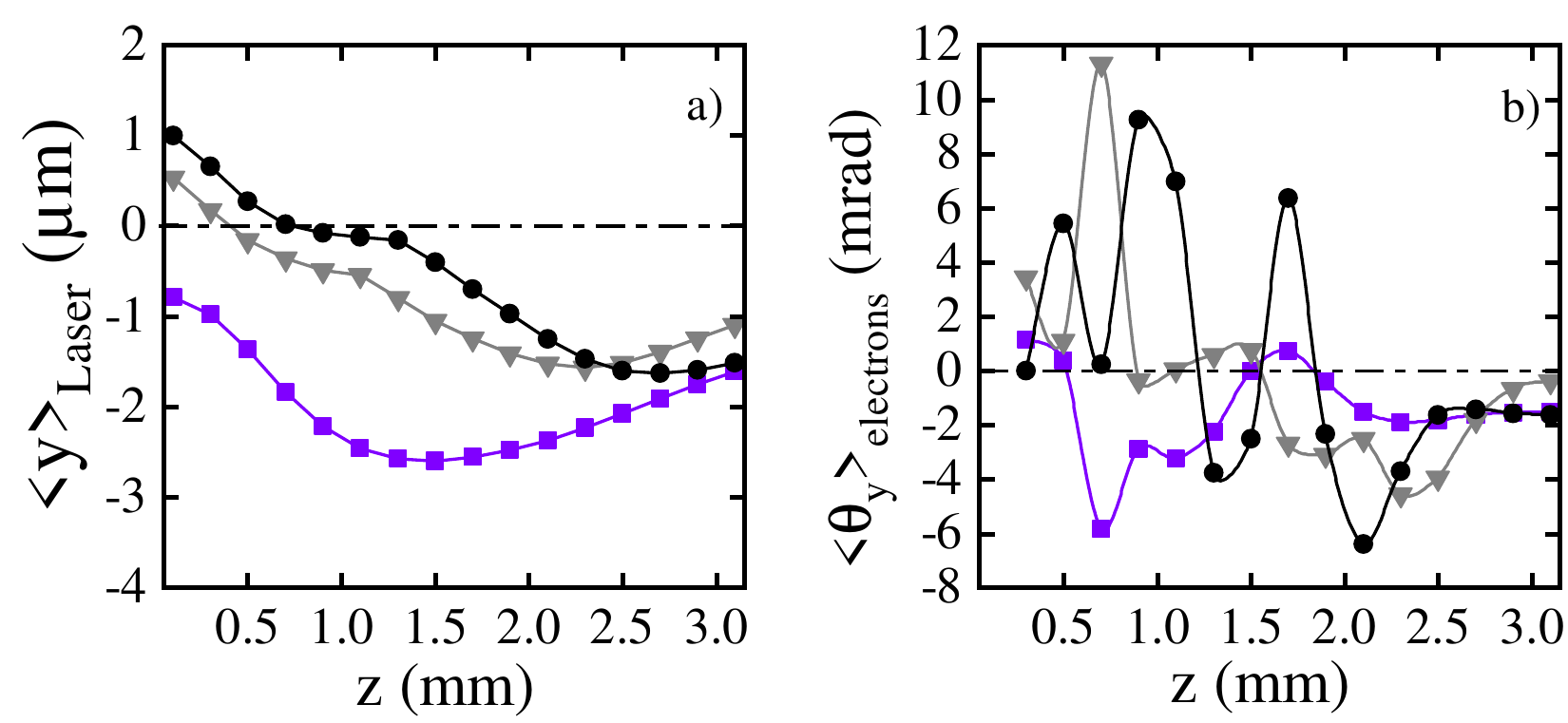}
\captionof{figure}{\label{fig:Fig7-LaserParticlesCorrelation} Simulation results for a) the evolution of the centre of mass of the laser fluence inside a disk of $20 \, \mu$m radius, and b) of the average angle of the electrons accelerated to the energy peak, as a function of position on the laser axis, for three focal positions: \SI{-0.8}{\milli\meter} (purple square symbols), \SI{+0.4}{\milli\meter} (grey triangles) and \SI{+0.7}{\milli\meter} (black circles).}
\end{figure}

In order to analyse more closely the correlation between the laser propagation and the transverse displacement of the electrons accelerated up to the peak energy, we have plotted in Fig. \ref{fig:Fig7-LaserParticlesCorrelation} the evolution of corresponding average values inside the plasma target with the same spatial units as Fig. \ref{fig:ShortVsLongDensity}. Here three cases are presented: a pre- and post-plateau focal position corresponding to positions a) and c) of Fig.  \ref{fig:Fig6_combined}, and a peri-plateau focal position corresponding to the simulated case with the lowest electron bunch axial deflection.

Simulation results for the evolution of the $y$ positions of the centre of mass of the laser fluence $<y>_{\rm{Laser}}$ during propagation, calculated over a transverse disk of $20 \, \mu$m radius centred on the z-axis, for three focal positions $z_{\mathrm{foc}}$ = \SI{-0.8}{\milli\meter}, \SI{+0.4}{\milli\meter} and \SI{+0.7}{\milli\meter} as parameters, are plotted in Fig. \ref{fig:Fig7-LaserParticlesCorrelation}a), corresponding to the case of 1bFPS AO settings, n$_e = 6.7\times 10^{18}$ cm$^{-3}$ and $a_0=2.15$ (blue squares in Fig.~6). Here we use $z_{\mathrm{foc}}$ to distinguish between the vacuum focal positions and the longitudinal position (z) dependent behaviour of the laser and electrons. During the first stage of propagation, $<y>_{\rm{Laser}}$ is decreasing for all focal positions, with an angular direction of the order of \SI{-1}{\milli\radian}, reflecting the asymmetry of the injected laser intensity profile. This decrease of $<y>_{\rm{Laser}}$ continues during self-focusing. This variation of $<y>_{\rm{Laser}}$ induces a displacement of the centre of mass of the accelerated electrons toward negative values of $y$. After $z=$\SI{1.5}{\milli\meter} for $z_{\mathrm{foc}}$ =\SI{-0.8}{\milli\meter}, $z=$\SI{2.5}{\milli\meter} for $z_{\mathrm{foc}} =$\SI{0.4}{\milli\meter} and $z_{\mathrm{foc}} =$ \SI{0.7}{\milli\meter}, the value of $<y>_{\rm{Laser}}$ either stabilises or increases, transverse diffraction of the laser becomes dominant with the decrease of plasma density.

Between $z=$\SI{0.5}{\milli\meter} and $z=$\SI{2}{\milli\meter}, the plasma wakefield has the highest amplitude, not only accelerating electrons but also producing transverse oscillations (so-called betatron oscillations) of the accelerated electrons as can be seen from the average electron angle evolution, plotted in Fig. \ref{fig:Fig7-LaserParticlesCorrelation}b). After $z$=\SI{2}{\milli\meter}, electrons can perform only a fraction of the betatron oscillation period, determining the final average angle at the exit. For $z_{\mathrm{foc}}$ = \SI{+0.4}{\milli\meter}, there is a nearly perfect final focusing, leading to a very small exit angle. At the same time, for the other two focal positions, the coupling between the laser intensity and density gradient have non-optimal values close to the plasma exit, resulting in larger final angles. 

It has already been reported that the exit gradient can be optimised to reduce the final RMS divergence of the electron bunch  \cite{mehrling2012transverse,Lehe2014_PhysRevSTAB,Li2019_PhysRevAB}. The interaction between the wake and electron bunch has been studied extensively in these references in terms of RMS bunch parameters. In addition we show that the asymmetry in the laser fluence profile can change the electron bunch axial displacement caused by the average bunch divergence as demonstrated in Fig \ref{fig:Fig7-LaserParticlesCorrelation}b). The magnitude of this effect can be controlled by modifying the laser-plasma coupling through a shift in the focal position.

In summary, the final angular deviation of the electron bunch is determined by three main factors: first, the initial symmetry of the focusing laser, second the position of the focal plane relative to the plasma density profile at which self-focusing becomes dominant, and third, laser amplitude and plasma density gradient at the exit region of the target. These results show that for optimal focal positions, the plasma density profile originating from the ELISA gas cell design, can efficiently reduce the angular deviation leading to better coaxiality of the electron bunch with the laser axis. Simulations show that this reduction of the angular deviation is efficient for both transverse directions, $x$ as well as $y$.

\subsection{Influence of Laser Wavefront on Electron Bunches}
\label{sec:ao_settings}

A third control mechanism was explored using the AO settings to study the influence of the laser wavefront on the injection process. This influence is analysed in more detail for the focal position $z_{\mathrm{foc}}$= \SI{-0.8}{\milli\meter} because, as seen in Fig. \ref{fig:Fig6_combined}a), it can produce electron bunches with single peak spectra and was not previously studied. In most previous works, either experimental or theoretical, the laser focal plane was set deep inside the plasma to optimise the position where the primary trapping process occurs \cite{couperus2017demonstration,Kirchen_PRL2021,lee2018optimization}. 

Results obtained at focal position $z=$\SI{-0.8}{\milli\meter} are compared in Fig.~\ref{fig:WF_allcases} for the three AO settings described in section I, FPS, 1bFPS and 2bFPS.
 
\begin{center}
\includegraphics[trim={0 0 0 0},clip, width=0.5\textwidth]{
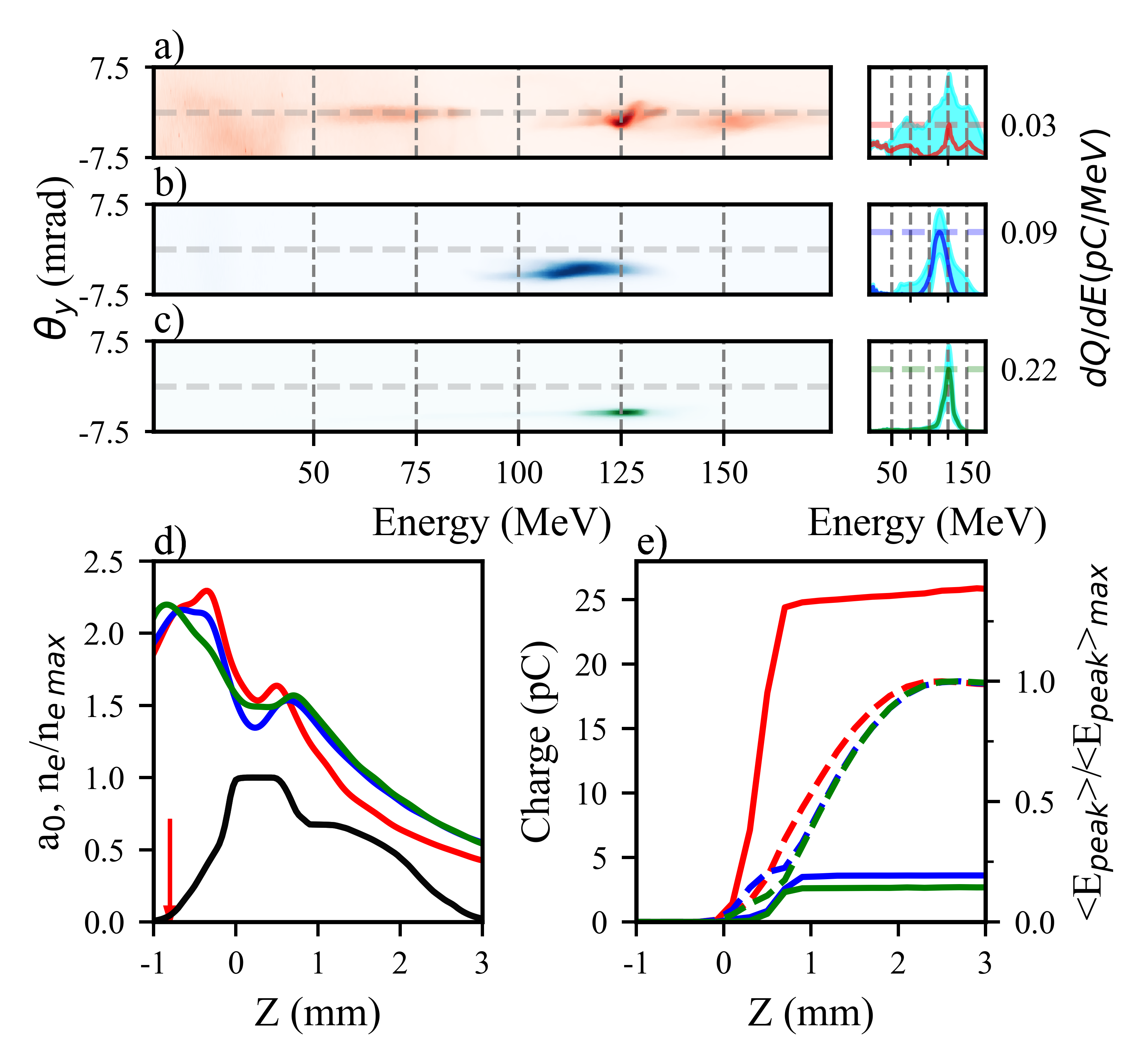}
\captionof{figure}{\label{fig:WF_allcases} Experimental spectra a)-c) illustrating the effect of phase front optimisation on accelerated electron bunches at laser focus $z_{\mathrm{foc}}$ = \SI{-0.8}{\milli\meter} for wavefront configurations a) FPS, b) 1bFPS and c) 2bFPS. Electron charge density in divergence-energy space (\si{\pico\coulomb\per\mega\electronvolt\per\milli\radian} and their corresponding spatially-integrated $\mathrm{dQ}/\mathrm{dE}$ (\si{\pico\coulomb\per\mega\electronvolt}) within a $\pm$\SI{7.5}{\milli\radian} window around the laser axis indicated by the dashed horizontal line and in an energy window \SI{11.3}{\mega\electronvolt} and \SI{175}{\mega\electronvolt}. Standard deviation calculated over five shots and plotted here in cyan. d) simulation results for the evolution along the propagation distance $z$ of the normalised vector potential $a_0$ of the laser pulse: the black curve represents the normalised plasma density profile, while the focus position, $z_{\mathrm{foc}}$, is marked by the red arrow; e) evolution with $z$ of the charge of the electrons having final energy above 10~MeV (solid lines) and average energy of the electrons contributing to the peak in energy normalised by its maximum values (dashed lines). For figures d-e, the red curves correspond to FPS, blue curves to 1bFPS, and green to 2bFPS.}
\end{center}
 
These three AO configurations (FPS, 1bFPS, 2bFPS) yield similar values for the total charge in the peak (1.6$\,\pm\,$0.3, 2.3$\,\pm\,$0.4, 3.7$\,\pm\,$0.7)~\si{\pico\coulomb} and for the peak energy (126$\,\pm\,$8, 114$\,\pm\,$7, 125$\,\pm\,$3)~\si{\mega\electronvolt}. However, the corresponding experimental electron spectra differ significantly for the FPS case, as seen from Fig.~\ref{fig:WF_allcases} a) to c).

Whereas FPS settings yield a broad spectrum in energy and a larger dispersion in angle, the 1bFPS and 2bFPS configurations generate single peaks with a lower dispersion both in energy (18\% and 8.7\% FWHM) and angles (4.4$\,\pm\,$0.6, 1.8$\,\pm\,$0.6)~\si{\milli\radian} full angle,  a maximum $\mathrm{dQ}/\mathrm{dE}$ of (0.09$\,\pm\,$0.02, 0.22$\,\pm\,$0.04) \si{\pico\coulomb\per\mega\electronvolt} and a bunch angular deviation of (-3.3$\,\pm\,$0.3, -4.3$\,\pm\,$0.3) \si{\milli\radian}; it contains  (93, 60)\% of the total charge in $\pm 1\times$ FWHM leading to a total energy of (0.2, 0.4)~\si{\milli\joule}. These values indicate that the 2bFPS configuration provides better quality electron bunches, with twice more energy in the peak together with a reduced dispersion in energy and angle. Simulation results are in good agreement with the experimental data.

To better understand the physics involved, we have plotted simulation results for the laser $a_0$ in Fig. \ref{fig:WF_allcases}d), and the evolution of the charge of the accelerated electrons and the average energy of the electrons contributing to the peak of energy in Fig. \ref{fig:WF_allcases}e). The three AO settings provide similar curves with two maxima for the evolution of $a_0$. The first maximum is due to the focusing of the incoming beam slightly increased by relativistic self-focusing of the front of the laser pulse. In contrast, the second maximum comes from the ponderomotive focusing of the rear of the pulse. 

Injection through ionisation occurs only if the laser field is high enough to tunnel ionise the ion N$^{5+}$, which occurs for $a_0 > 1.5$. Fig.~\ref{fig:WF_allcases}d) shows this corresponds to the zone around the first ($z\approx-0.5$~mm) and second maxima ($z\approx1$~mm). Once generated through N$^{5+}$ ionisation, an electron needs also to be trapped by the plasma-wave field. Trapping requires a high enough plasma density, moreover, it is greatly favoured by a rapid increase of the longitudinal length of the positively charged bubble just behind the laser pulse. This increase occurs either in a density downramp or by a rapid increase of the laser intensity. Figure~\ref{fig:WF_allcases}e) shows that trapping occurs at the position of the second maximum for cases 1bFPS and 2bFPS. The slight increase in the value of $a_0$ for FPS causes trapping for $0<z<0.5$~mm, and a small amount of trapping throughout the downramp, as shown by the increase in total charge leading to the stronger low energy electron signal of Fig.~\ref{fig:WF_allcases}a). For 1bFPS and 2bFPS, no trapping occurs around the first maximum of $a_0$ because either the density is increasing or the intensity is decreasing. On the other hand, the zone $0.5<z<1.0$~mm around the second maximum is optimised for trapping: the density is decreasing and the intensity is increasing. Comparison of $a_0$ curves for FPS and 1bFPS configurations shows that in the FPS case, the second maximum is slightly higher and at a slightly smaller value of $z$. As $a_0$ values are close to the threshold $a_0 = 1.5$, small variations of $a_0$ result in a large difference in the trapped charge and in the energy spectra. In particular, the higher value of $a_0$ observed for the FPS results in a larger trapping zone, thus producing a broader energy spectrum. This high sensitivity at $z_{\mathrm{foc}}$=\SI{-0.8}{\milli\meter} also explains the fact that the total charge obtained in simulation for FPS can be higher than the experimental value.

The average energy of the peak electrons has a similar behaviour for the three AO settings, increasing up to the plasma exit and showing that the acceleration distance is smaller than the dephasing length. 
For $z_{\mathrm{foc}}$ =\SI{-0.8}{\milli\meter}, trapping occurs at low densities, putting the electrons at a large distance behind the laser pulse, therefore increasing the length of acceleration compared to trapping at positions close to $z_{\mathrm{foc}} = 0$.    

1bFPS and 2bFPS settings lead to very similar results, particularly concerning the evolution of the laser amplitude $a_0$ in Fig. \ref{fig:WF_allcases}d). In terms of electron trapping, the main difference is that the second 2bFPS peak is localised at a slightly larger $z$ than the 1bFPS one. As a consequence of this small shift, trapping of electrons starts slightly later for 2bFPS (at a lower density) and has a smaller duration, leading to a reduction of the energy spread and a small increase in the peak energy, because, in 2bFPS, the electrons are localised at a slightly larger distance from the laser pulse. As pointed out previously, close to the ionisation threshold of $N^{5+}$ in the trapping zone, the total charge is strongly dependent on the exact position of the focal plane. Nevertheless, the acceleration process depends only weakly on the total charge in the regime achieved here, where beam-loading does not contribute significantly.

These electrons bunches are deflected from the laser axis by approximately \SI{4}{\milli\radian} and additional mechanisms must be introduced to keep accelerated bunches on-axis whilst retaining high bunch quality. In comparison to the results presented in Fig. \ref{fig:Fig6_combined} it could be assumed that the target beam parameters (\SI{150}{\mega\electronvolt}, 5\% energy spread, \SI{30}{\pico\coulomb}) could be achieved through simply improving the laser quality closer to that of a Gaussian beam as at $z_{\mathrm{foc}}$ =\SI{-0.8}{\milli\meter} this provides \SI{50}{\pico\coulomb}. However, in this configuration the injection volume is increased due to the longer distance over which $a_0$ exceeds the injection threshold, leading to larger energy spread of the resulting spectra. This effect is seen in Fig. \ref{fig:WF_allcases}a) where FPS, the most symmetric setting (Fig. \ref{fig:LaserRotSymm}a)), produces the broadest spectra in energy and stronger fluctuations in consecutive electron spectra as illustrated by the standard deviation in the $\mathrm{dQ}/\mathrm{dE}$ plot. Optimisation of the LPI therefore requires the simultaneous tuning of a larger number of experimental parameters. The rotational symmetry of the laser pulse, as discussed in section \ref{LaserProperties}, could be used as an input parameter for an optimisation scheme using, for example, a Bayesian optimisation model \cite{shalloo2020automation, jalas2021bayesian} to produce electron bunches with the target parameters using the large experimental parameter space. Further, the use of pulse rotational symmetry would provide a simple input parameter in an optimisation model allowing for a reduction in the complexity usually associated with controlling individual AO pistons, or their corresponding Zernike polynomials. 

\section{Conclusions}

Previously predicted improvements in injector electrons with an elongated plasma density downramp are realised for the ELISA gas cell \cite{Lee2016,lee2018optimization}. Advances in the understanding of LPI are achieved through experimental and simulated studies of the laser-plasma coupling with alterations in plasma density structure, focal position, and laser pre-focal symmetry.

To optimise the injector, we have selected three main parameters for their significant impact on the resulting electron spectra in the regime studied: the length of the plasma density downramp to control the acceleration and focusing fields which the trapped electrons experience, the focal position to control the non-linear coupling between the laser and the plasma, and the laser wavefront to alter the transverse energy distribution of the laser through focus to control the dynamics of the wakefield through the effect of self-focusing. Through careful optimisation of the density downramp, focal position, and shaping of laser symmetry, electron bunches with energy in the 100~MeV range, less than 10 per cent energy spread, multi-\si{\pico\coulomb} charge and sub \SI{2}{\milli\radian} divergence are produced as illustrated in Fig. \ref{fig:WF_allcases}c). Comparison to realistic simulations uncovers the physical mechanisms controlling the electron dynamics which produce these desirable bunches. Bunch energy (\SI{125}{\mega\electronvolt}) and energy spread (8.7\%) approach the desired values (\SI{150}{\mega\electronvolt}, 5\%), however the level of charge must be augmented significantly by a factor of 8 from \SI{3.7}{\pico\coulomb} to \SI{30}{\pico\coulomb}, to reach the desired value for an LPI within the EuPRAXIA framework. Future optimisation should explore larger parameter spaces to ameliorate the results of this novel injection mechanism using the rotational symmetry as an input parameter. 

Changing the pre-focal symmetry whilst retaining a similar focal spot is shown to have a measurable effect on the accelerated electron bunches and suggests that this could be another control mechanism to utilise when optimising LPIs. Simulations using realistic laser parameters produced accurate descriptions of the accelerated bunch dynamics whilst Gaussian models failed to achieve this. This work expands on the physics of injectors and provides a simulation method using realistic laser parameters for improving the accuracy of predictions for laser wakefield acceleration schemes in a computationally inexpensive way.

\section*{Acknowledgments}
This project has received funding from the European Union’s Horizon 2020 Research and Innovation Programme under Grant Agreement No. 730871. We acknowledge support from the Knut and Alice Wallenberg Foundation (Grant No. KAW 2019.0318) and the Swedish Research Council (Grant No. 2019-04784). We acknowledge the computing center MesoLUM managed by Institut des Sciences Moléculaires d'Orsay (UMR8214) and Laboratoire de Physique des Gaz et des Plasmas (UMR8578), University Paris-Saclay (France).

\newpage

\bibliography{bib}

\newpage

\end{document}